\documentclass[final,3p,times]{elsarticle}

\usepackage{amssymb}

\journal{Signal Processing}

\usepackage{color}
\usepackage{graphicx}  
\usepackage{subfigure} 
\usepackage{multirow}

\newcommand{\e}{\begin{equation}}
\newcommand{\ee}{\end{equation}}
\newcommand{\eqn}{\begin{eqnarray}}
\newcommand{\eeqn}{\end{eqnarray}}
\newcommand{\MAT}{\left[ \begin{array}}
\newcommand{\mat}{\end{array} \right]}
\newcommand{\pl}{\left(}
\newcommand{\pr}{\right)}
\newcommand{\bl}{\left[}
\newcommand{\br}{\right]}

\newcommand{\dvl}{\left\|}
\newcommand{\dvr}{\right\|}

\newcommand{\calL}{\mathcal{L}}
\newcommand{\calT}{\mathcal{T}}

\newcommand{\setR}{\mathbb{R}}

\newcommand{\vct}[1]{\boldsymbol{#1}}
\newcommand{\mtx}[1]{\boldsymbol{#1}}

\newcommand{\ve}{\vct{e}}

\newcommand{\vx}{\vct{x}}
\newcommand{\vy}{\vct{y}}
\newcommand{\vz}{\vct{z}}

\newcommand{\vepsilon}{\vct{\epsilon}}

\newcommand{\vlambda}{\vct{\lambda}}

\newcommand{\vxi}{\vct{\xi}}

\newcommand{\mA}{\mtx{A}}

\newcommand{\mI}{\mtx{I}}

\newcommand{\mU}{\mtx{U}}
\newcommand{\mV}{\mtx{V}}

\newcommand{\mX}{\mtx{X}}

\newcommand{\mSigma}{\mtx{\Sigma}}

\usepackage{bm}   
\usepackage{amsmath}  
\usepackage{url}
\usepackage{amsfonts}  
\usepackage{lscape}

\begin{document}

\begin{frontmatter}



\title{PIPO-Net: A Penalty-based Independent Parameters Optimization Deep Unfolding Network}


\author{Xiumei Li$^1$, Zhijie Zhang$^1$, Huang Bai$^{1\ast}$, Ljubi\v{s}a Stankovi\'{c}$^2$, Junpeng Hao$^1$, and Junmei Sun$^1$}

\address{$^1$School of Information Science and Technology, Hangzhou Normal University, Hangzhou, 311121, Zhejiang, China\\
$^2$Faculty of Electrical Engineering, University of Montenegro, Podgorica, Montenegro}

\begin{abstract}
Compressive sensing (CS) has been widely applied in signal and image processing fields.
Traditional CS reconstruction algorithms have a complete theoretical foundation but suffer from the high computational complexity,
while fashionable deep network-based methods can achieve high-accuracy reconstruction of CS but are short of interpretability.
These facts motivate us to develop a deep unfolding network named the penalty-based independent parameters optimization network (PIPO-Net)
to combine the merits of the above mentioned two kinds of CS methods.
Each module of PIPO-Net can be viewed separately as an optimization problem with respective penalty function.
The main characteristic of PIPO-Net is that, in each round of training,
the learnable parameters in one module are updated independently from those of other modules.
This makes the network more flexible to find the optimal solutions of the corresponding problems.
Moreover, the mean-subtraction sampling and the high-frequency complementary blocks are developed to improve the performance of PIPO-Net.
Experiments on reconstructing CS images demonstrate the effectiveness of the proposed PIPO-Net.
\end{abstract}


\end{frontmatter}


\section{Introduction}\label{section1}
Compressive sensing (CS), as a novel signal processing scheme,
intends to reconstruct some high-dimensional signals by using much fewer samples than that required by the Nyquist sampling theorem.
Based on its rich theoretical analysis results, CS has attracted a lot of attention in signal processing society \cite{PMH23}, \cite{UGC23}.

Traditional CS reconstruction algorithms utilize the structured sparsity of the interested signals
as the prior knowledge to solve sparse regularized optimization problems in an iterative manner \cite{KNB10} - \cite{LBH17},
e.g., the iterative hard thresholding algorithm (IHT) \cite{BD08} and the iterative soft thresholding algorithm (ISTA) \cite{DDDM04}.
Donoho {\em et al.} combined ISTA with the Onsager to achieve an efficient algorithm called approximate message passing (AMP) \cite{DMM09}.
However, many real signals, such as images, are compressible only in certain transform domains, but are not exact sparse.
The reconstruction accuracy of such signals is low when handcrafted sparse prior knowledge is utilized for iteration.

With the development of artificial intelligence technology, deep learning is widely applied
in many fields such as image classification, super-resolution and natural language processing.
Inspired by the super-resolution network, Kulkarni {\em et al.} proposed the ReconNet to reconstruct CS images
using a convolutional neural network (CNN) based approach \cite{KLTKA16}.
Shi {\em et al.} proposed an image CS framework CSNet using CNN that included
a sampling network and a reconstruction network, which were optimized jointly \cite{SJLZ20}.
These deep network-based non-iterative algorithms can significantly reduce the time complexity compared to traditional CS reconstruction algorithms.
However, these networks consist of repeated fully connected (FC) layers or convolutional layers,
which are essentially black boxes without good interpretability and theoretical guarantee.

As a new class of deep learning models, deep unfolding networks can fuse the
interpretability of traditional iterative reconstruction algorithms and
the high performance of classical deep network-based models
to quickly achieve high-accuracy reconstruction of CS \cite{YSLX20} - \cite{ZLLWZ21}.
Khobahi {\em et al.} investigated the problem of one-bit CS and proposed a novel hybrid
autoencoding scheme that allowed to jointly learn the parameters of the measurement module
and the latent-variables of the decoder function \cite{KS20}.
Zeng {\em et al.} presented a novel deep architecture which was able to learn the sampling matrix
by taking advantage of the unfolded algorithm such that the recovery algorithm can quickly
recover the signal from its one-bit noisy measurements \cite{ZKS22}.
Zhang {\em et al.} proposed ISTA-Net and ISTA-Net$^+$,
which used a symmetrically structured CNN to learn the optimal transform domain \cite{ZG18}.
Su {\em et al.} developed a nonconvex optimization inspired multi-scale CS reconstruction network,
iPiano-Net, by unfolding the classic iPiano algorithm \cite{SL20}.
Zhou  {\em et al.} proposed a multi-channel deep network termed BCSnet for block-based image CS by exploiting inter-block correlation
based on the popular BCS-SPL algorithm \cite{ZHLLZ21}.
Zhang {\em et al.} built the AMP-Net based on the denoising perspective of the traditional AMP algorithm \cite{ZLLWZ21}.
Generally, a deep unfolding-based reconstruction network consists of multiple modules,
and each module corresponds to one iteration of the iterative reconstruction algorithm.

In this paper, a penalty-based independent parameters optimization deep unfolding network (PIPO-Net) is proposed
to achieve fast and high-accuracy reconstruction of image CS.
Different from previous networks where per module strictly corresponds to one iteration of the iterative algorithm,
each module of our proposed PIPO-Net can be viewed separately as an optimization problem
to be solved by a CS model with respective penalty function.
Therefore, the problem of each module is relatively independent from the results of other modules,
and the solution would not easily fall into the local minimum.
Experiments show that PIPO-Net outperforms classical deep network-based CS reconstruction methods including the deep unfolding network-based ones.

The remainder of this paper is arranged as follows.
In Section~\ref{section2}, the detailed architecture of PIPO-Net is described with each part well explained.
Experiments are carried out in Section~\ref{section3} to verify the performance of the proposed network.
Some concluding remarks are given in Section~\ref{section4} to end this paper.

\section{PIPO-Net}\label{section2}
This paper investigates the generic penalty-based CS reconstruction model:
\e \arg\min_{\vx}~\frac{1}{2}\dvl\vy-\mA\vx\dvr_2^2+\omega g(\vx),\label{penalty-CS} \ee
where $\vx\in\setR^{n\times 1}$ denotes the original signal vector, $\mA\in\setR^{m\times n}$ with $m\ll n$ is the CS sampling matrix,
$\vy\in\setR^{m\times 1}$ is the measurement vector, and $g(\vx)$ means the penalty function with $\omega$ the penalty parameter.
Most traditional CS reconstruction methods set $g(\vx)$ to be the $\ell_0$-norm or $\ell_1$-norm constraint on signal $\vx$ in some transform domain,
which has a good theoretical analysis, but encounters issues such as high computational complexity,
difficult selections on optimal transformation and parameters \cite{SSSDO19}.
In this paper, the powerful learning ability of deep network is utilized to learn the features of $\vx$,
and set $g(\vx)$ as an adaptive function to be optimized and solved through the network.

In order to separate the data fidelity term and the penalty term in (\ref{penalty-CS}), an auxiliary variable $\vz$ is introduced such that:
\e \arg\min_{\vx}~\frac{1}{2}\dvl\vy-\mA\vx\dvr_2^2+\omega g(\vz),~~~\text{s.t.}~~~\vz=\vx.\label{penalty-zx} \ee
The augmented Lagrange function can be constructed:
\e {\mathrm{L}}_\rho(\vx,\vz,\vlambda)~\triangleq~\frac{1}{2}\dvl\vy-\mA\vx\dvr_2^2+\omega g(\vz)
+\vlambda^{\calT}\pl\vz-\vx\pr+\frac{\rho}{2}\dvl\vz-\vx\dvr_2^2,\label{ALfun} \ee
where $\vlambda\in\setR^{n\times 1}$ is the Lagrange multiplier vector, $\calT$ is the transpose operator,
and $\rho$ is a proper penalty parameter. For the above multivariate optimization problem,
the following alternating optimization over $\vz$, $\vlambda$ and $\vx$ can be conducted to reach the solutions:
\e
\left\{\begin{array}{ll} \arg\min_{\vz}~\frac{\omega}{\rho}g(\vz)+\frac{1}{2}\dvl\vz-\pl\frac{\vlambda}{\rho}+\vx\pr\dvr_2^2,\\
\vlambda~\leftarrow~\vlambda+\rho\pl\vz-\vx\pr,\\
\arg\min_{\vx}~\frac{1}{2}\dvl\vy-\mA\vx\dvr_2^2+\vlambda^{\calT}\pl\vz-\vx\pr+\frac{\rho}{2}\dvl\vz-\vx\dvr_2^2.
\end{array} \right. \label{sub-problems}\ee

To avoid the difficulties in obtaining the optimal transformation and parameters,
PIPO-Net is proposed, which combines deep networks with traditional algorithms to learn the above parameters from data through the network.

\subsection{PIPO-Net architecture}
\begin{landscape}
\begin{figure}[htb!]
\centering
\includegraphics[width=8.5in]{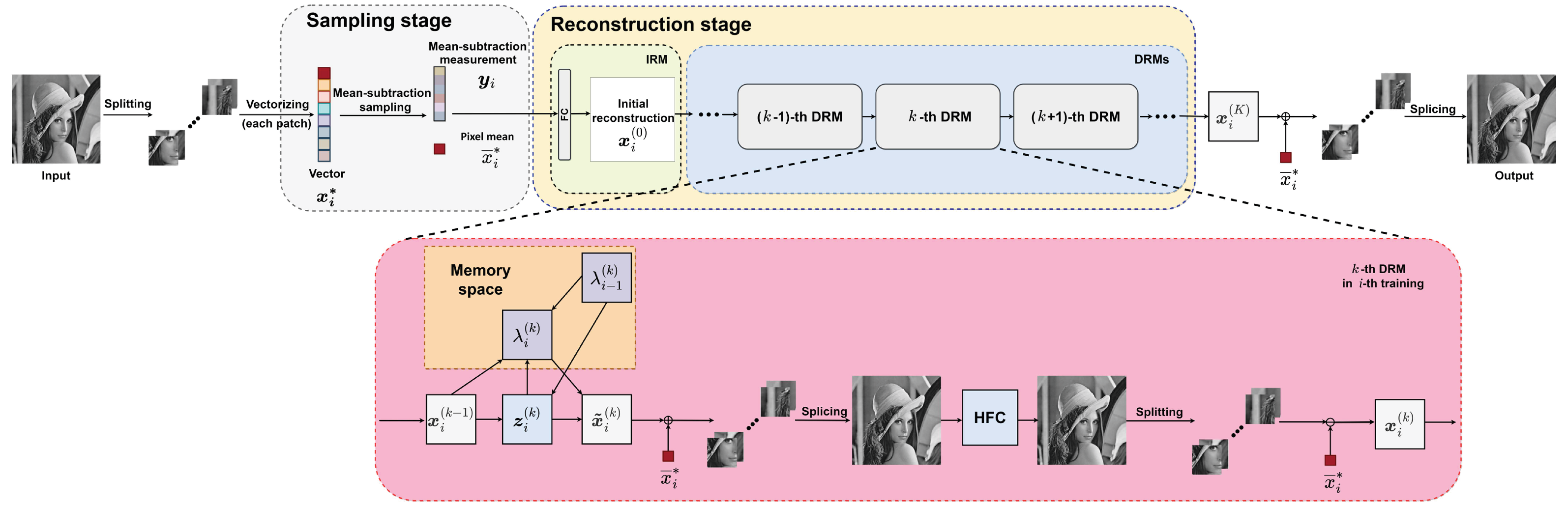}
\caption{PIPO-Net architecture.}\label{PIPONetwork}
\end{figure}
\end{landscape}

The popular image patch-based training strategy \cite{ZG18} - \cite{ZLLWZ21} is adopted for the proposed PIPO-Net.
The training is carried out for several rounds.
Numerous whole images are split into non-overlapping patches of size $\sqrt{n}\times\sqrt{n}$ and the patches extracted from the same image should be
utilized in the same round of training.
For the $i$-th (round of) training, the architecture of PIPO-Net is described in Fig.~\ref{PIPONetwork}.

The network architecture mainly consists of the sampling stage and the reconstruction stage.
In the sampling stage, each input patch is firstly rearranged into the training vector $\vx_i^\ast$ of size $n\times 1$.
A strategy named mean-subtraction sampling (MSS) is designed to realize the compressive sampling of $\vx_i^\ast$ with its pixel mean removed.
The outputs of this sampling stage include the mean-subtraction measurement $\vy_i$ and the patch pixel mean $\bar{x}_i^\ast$.

The reconstruction stage contains an initial reconstruction module (IRM) and $K$ deep reconstruction modules (DRMs).
With the mean-subtraction measurement $\vy_i$, the initial reconstruction $\vx_i^{(0)}$ is achieved by employing the FC layer.
By transmitting $\vx_i^{(0)}$ pass through $K$ DRMs in sequence, high-accuracy reconstruction would be obtained.
Each DRM mainly consists of three variable update blocks that are: $\vz$-update block, $\vlambda$-update block and $\vx$-update block.
In the $k$-th DRM, the auxiliary variable $\vz_i^{(k)}$ is firstly updated by using $\vx_i^{(k-1)}$ outputted from the $(k-1)$-th DRM
and $\vlambda_{i-1}^{(k)}$ stored in the $(i-1)$-th training.
Then the multiplier vector $\vlambda_{i}^{(k)}$ can be updated by taking $\vlambda_{i-1}^{(k)}$, $\vx_i^{(k-1)}$ and $\vz_i^{(k)}$,
and $\vlambda_{i-1}^{(k)}$ in the memory would be overwritten by the updated $\vlambda_{i}^{(k)}$.
With the obtained $\vz_i^{(k)}$ and $\vlambda_{i}^{(k)}$, the basic image signal $\tilde{\vx}_i^{(k)}$ is updated.
In order to eliminate the blocking artifacts and recover the missing high-frequency information,
an additional high-frequency complementary (HFC) block is incorporated at the end of each DRM.
By utilizing the basic image signal $\tilde{\vx}_i^{(k)}$ and the patch pixel mean $\bar{x}_i^\ast$,
each basic image patch could be recovered to splice the whole image to enter the HFC block,
where a data-driven sub-network is developed to learn the missing high-frequency information.
The output image of HFC is then once again split into patches and $\vx_i^{(k)}$ is achieved for the next DRM.

As can be observed that only the image signal is transferred between modules,
while parameters in one DRM are updated independently from those of other DRMs.
The designs of the MSS, the three variable update blocks,
and the HFC block will be detailedly introduced in the following subsections.

\subsection{Mean-subtraction sampling}
Mean-subtraction data, also known as zero-centered data, play important roles in signal processing tasks.
For the training of the network, removing the pixel mean in the image would
speed up the convergence of the parameters at each layer, thus accelerating the CS reconstruction.

CS aims to sample and compress the signal simultaneously,
and it is not possible to directly sample the mean-subtraction image data without the prior knowledge on the image.
Here, an MSS strategy is designed to get the mean-subtraction measurement
while slightly increasing the sampling complexity.

Assume that $\vx^{\ast}=[x_1,x_2,\cdots,x_j,\cdots,x_n]^{\calT}\in\setR^{n\times 1}$.
The patch pixel mean $\bar{x}^{\ast}=1/n\sum_{j=1}^n x_j$.
Define a pixel mean vector $\bar{\vx}=[\bar{x}^\ast,\bar{x}^\ast,\cdots,\bar{x}^\ast]^{\calT}\in\setR^{n\times 1}$.
Let the mean-subtraction image data $\vx=\vx^\ast-\bar{\vx}$. With sampling matrix $\mA\in\setR^{m\times n}$,
the MSS process can be formulated as $\vy=\mA\vx$.
Since there is no prior knowledge on $\vx^{\ast}$,
the following augmented sampling matrix is constructed:
\e \mA^\ast~\triangleq~\MAT{ccc}~&\mA&~\\1&\cdots&1\mat~\in~\setR^{(m+1)\times n}.\label{ASM} \ee
Therefore, the actual sampling process is
\e \vy^\ast~=~\mA^\ast\vx^\ast~=~\MAT{cc}\mA\vx^\ast\\ n\bar{x}^\ast\mat~\triangleq~\MAT{cc}\tilde{\vy}\\ y_{m+1}\mat, \ee
where $\tilde{\vy}\triangleq\mA\vx^\ast$ and $y_{m+1}\triangleq n\bar{x}^\ast$. Hence, the mean-subtraction measurement can be obtained as:
\e \vy~=~\mA\vx~=~\mA(\vx^\ast-\bar{\vx})~=~\mA\vx^\ast-\mA\bar{\vx}
~=~\tilde{\vy}-\frac{1}{n}\mA\bl~y_{m+1},y_{m+1},\cdots,y_{m+1}~\br^{\calT}. \ee
By introducing the augmented sampling matrix, the MSS of the image is realized.

Since the sampling matrix plays an important role in both sampling and reconstruction in most deep unfolding methods,
a properly designed sampling matrix can effectively enhance the reconstruction performance \cite{ZLLWZ21}.
The same gradient-based sampling matrix training strategy is adopted in the proposed network as in AMP-Net \cite{ZLLWZ21},
where the sampling matrix is jointly trained with other network parameters to improve the reconstruction accuracy.
For the gradient-based method, a carefully chosen initialization is crucial for ensuring convergence.
In the following, we introduce an initial sampling matrix chosen scheme according to the maximum likelihood estimation (MLE) principle.

Suppose that the original signal $\vx$ is interfered by noise $\ve$ which follows a normal distribution ${\cal N}\pl{\bm 0},\nu\mI_n\pr$
where $\mI_n$ denotes the identity matrix of size $n\times n$ and $\nu$ is the variance. Then the measurement can be expressed as:
\e \vy~=~\mA\pl\vx+\ve\pr~=~\mA\vx+\mA\ve~\triangleq~\mA\vx+\vepsilon. \ee
According to the distribution of $\ve$, its probability density function (PDF) is
\e f_{\ve}\pl\vxi\pr~=~\frac{e^{-\frac{1}{2}\vxi^{\calT}\pl\nu\mI_n\pr^{-1}\vxi}}{\sqrt{2\pi\text{det}\pl\nu\mI_n\pr}}, \ee
in which $\text{det}(\cdot)$ denotes the determinant operator.
$\vepsilon\triangleq\mA\ve$ indicates that $\vepsilon$ has a multivariate normal distribution ${\cal N}\pl{\bm 0},\nu\mA\mA^{\calT}\pr$,
which means that the PDF of $\vepsilon$ obeys
\e f_{\vepsilon}\pl\vxi\pr~=~\frac{e^{-\frac{1}{2}\vxi^{\calT}\pl\nu\mA\mA^{\calT}\pr^{-1}\vxi}}{\sqrt{2\pi\text{det}\pl\nu\mA\mA^{\calT}\pr}}. \ee
According to the MLE principle \cite{Kay93}, the best estimate of $\vx$ that can be recovered from the measurement $\vy$
is the one that maximizes the likelihood function on $\vx$, that is, the PDF $f_{\vepsilon}\pl\vxi\pr$ at $\vxi=\vy-\mA\vx$,
which leads to
\e \max_{\vx}~f_{\vepsilon}\pl\vxi\pr~~\Leftrightarrow~~\max_{\vx}~e^{-\frac{1}{2}\vxi^{\calT}\pl\nu\mA\mA^{\calT}\pr^{-1}\vxi}
~~\Leftrightarrow~~\min_{\vx}~\vxi^{\calT}\pl\nu\mA\mA^{\calT}\pr^{-1}\vxi
~~\Leftrightarrow~~\min_{\vx}~\dvl\pl\mA\mA^{\calT}\pr^{-1/2}\pl\vy-\mA\vx\pr\dvr_2^2,\label{modified-CS} \ee
as from a statistical point of view, the variance $\nu\geq 0$ is independent from $\vx$.

When the noise $\ve$ is not null, comparing the last term in (\ref{modified-CS}) with the data fidelity term in (\ref{penalty-CS}),
the MLE principle indicates that a sampling matrix $\check{\mA}=\pl\mA\mA^{\calT}\pr^{-1/2}\mA$ can enhance the signal reconstruction performance.
Assume that a general full row rank matrix $\mA\in\setR^{m\times n}$ has the following singular value decomposition (SVD):
\e \mA~=~\mU\MAT{cc}\mSigma& {\bm 0}\mat\mV^{\calT} \ee
with $\mU$ and $\mV$ being two orthonormal matrices of proper dimensions, and $\mSigma$ being the diagonal singular value matrix.
It can be shown that
\e \check{\mA}~=~\pl\mA\mA^{\calT}\pr^{-1/2}\mA~=~\mU\MAT{cc}\mI_m& {\bm 0}\mat\mV^{\calT},\label{SM-form} \ee
as $\pl\mA\mA^{\calT}\pr^{-1/2}=\mU\mSigma^{-1}\mU^{\calT}$.

Based on the discussions above, the sampling matrix will be initialized under the form (\ref{SM-form}).
In practice, such a sampling matrix can be obtained by randomly selecting a number of rows from an orthonormal matrix.

\subsection{Deep reconstruction module}
This subsection is devoted to the designs of the blocks in each DRM.
The $k$-th DRM in the $i$-th training will be taken as the example.

\subsubsection{$\vz$-update block}
\begin{figure}[htb!]
\centering
\includegraphics[width=6.25in]{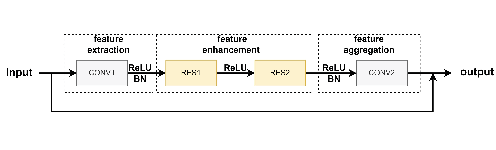}
\caption{$\text{Net}_z$ architecture.}\label{ZNetwork}
\end{figure}

The sub-network for updating $\vz$ is denoted as $\text{Net}_z$, which can be viewed as a nonlinear proximal mapping operator of $g(\vz)$.
For computational simplicity, $\omega=\rho$ is always set. The model of $\text{Net}_z$ is described as:
\e \vz_i^{(k)}~=~\text{Net}_z\pl\frac{\vlambda_{i-1}^{(k)}}{\rho_i^{(k)}}+\vx_i^{(k-1)}\pr,\label{ALfun-z-net} \ee
where $\rho_i^{(k)}$ is set as a learnable parameter in the current DRM and is independent from those of other DRMs.
The architecture of $\text{Net}_z$ is shown in Fig.~\ref{ZNetwork}.

This sub-network contains three parts that are: feature extraction, feature enhancement and feature aggregation.
The extraction part is composed of a convolution layer CONV1, batch normalization (BN) and rectified linear unit (ReLU),
aiming at extracting the features of the input data.
The enhancement part consists of two residual blocks RES1 and RES2 with ReLU between them.
This part is devoted to further extracting and reinforcing the features that have been captured in the previous part,
and it helps to the stepwise extraction of deeper feature representations for the follow-up part.
The aggregation part contains BN, ReLU and a convolution layer CONV2,
intending to integrate the obtained features to better reveal the complex features of the input data.
The convergence of the sub-network can be facilitated by fusing the inputs and outputs together through the skip connections.

\subsubsection{$\vlambda$-update block}
Using $\vlambda_{i-1}^{(k)}$ stored in the $(i-1)$-th training, $\vx_i^{(k-1)}$ outputted from the $(k-1)$-th DRM and the updated $\vz_i^{(k)}$,
$\vlambda_{i}^{(k)}$ is updated via
\e \vlambda_{i}^{(k)}~=~\vlambda_{i-1}^{(k)}+\rho_i^{(k)}\pl\vz_i^{(k)}-\vx_i^{(k-1)}\pr.\label{ALfun-lambda-form} \ee
Note that $\vlambda_{i-1}^{(k)}$ stored in the memory would be overwritten by the updated $\vlambda_{i}^{(k)}$.

\subsubsection{$\vx$-update block}
With the results of $\vz$-update block and $\vlambda$-update block,
the $\vx$-update problem in (\ref{sub-problems}) is solved by setting the derivative to zero,
which results in the basic image signal as
\e \tilde{\vx}_i^{(k)}=\pl\mA^{\calT}\mA+\rho_i^{(k)}\mI_n\pr^{-1}
\pl\mA^{\calT}\vy_i+\vlambda_{i}^{(k)}+\rho_i^{(k)}\vz_i^{(k)}\pr,\label{ALfun-x-form} \ee
where $\mA$ is the sampling matrix, $\vy_i$ is the mean-subtraction measurement in the $i$-th training,
and $\mI_n$ is the identity matrix of size $n\times n$.

\subsubsection{HFC block}

The proposed PIPO-Net adopts the popular image patch-based training strategy \cite{ZG18} - \cite{ZLLWZ21}.
One image is measured patch-by-patch using the same sampling matrix.
However, splicing the recovered image patches into the whole image usually leads to undesired blocking artifacts.
When additional deblocking operations are executed to eliminate these artifacts, some internal high-frequency information would be lost.
Hence, the HFC block is designed in each DRM to compensate the missing high-frequency information and remove the annoying artifacts.
The same architecture as $\text{Net}_z$ is adopted for the HFC block,
denoted as $\text{Net}_{\theta_i^{(k)}}$ with $\theta_i^{(k)}$ the parameters to be learned.
The input of $\text{Net}_z$ is the sequence of the vectorized image patches,
while $\text{Net}_{\theta_i^{(k)}}$ deals with the whole image.

To enforce the HFC blocks to learn the high-frequency information, a two-dimensional discrete wavelet transform (DWT) is employed
to decompose the whole image into four sub-band coefficients, and the following wavelet loss function is developed:
\e \calL_{\text{WT}}~\triangleq~\frac{1}{NK}\sum_{j=1}^{N}\sum_{k=1}^{K}
\dvl\text{WT}\pl\mX_j\pr-\text{WT}\pl\text{Net}_{\theta_i^{(k)}}\pl\tilde{\mX}_j^{(k)}\pr\pr\dvr_F^2, \ee
where $\text{WT}(\cdot)$ denotes the Haar wavelet decomposition operation,
$\|\cdot\|_F$ is the Frobenius norm,
$K$ is the number of HFC blocks (i.e., the number of DRMs),
$N$ is the number of whole images chosen for the $i$-th training.
$\mX_j$ is one original whole image
and $\tilde{\mX}_j^{(k)}$ is the corresponding whole image as the HFC input in the $k$-th DRM.

Combining $\calL_{\text{WT}}$ with the commonly adopted mean square error (MSE) loss function $\calL_{\text{MSE}}$ defined as
\e \calL_{\text{MSE}}~\triangleq~\frac{1}{\text{numel}\pl\mX_j\pr N}\sum_{j=1}^{N}
\dvl\mX_j-\hat{\mX}_j\dvr_F^2, \ee
with $\text{numel}(\cdot)$ counting the number of pixels in one image
and $\hat{\mX}_j$ being the corresponding output image of the network,
the total loss function of PIPO-Net is presented as
\e \calL_{\text{total}}~=~\calL_{\text{MSE}}+\gamma\calL_{\text{WT}}, \ee
where $\gamma$ is the regularization parameter and $\gamma=0.01$ leads to satisfactory results.

\section{Experimental results}\label{section3}
In this section, the proposed PIPO-Net is applied for reconstruction of CS images.
Two classical deep network-based methods, ReconNet \cite{KLTKA16}, CSNet \cite{SJLZ20},
and four deep unfolding network-based methods, ISTA-Net$^+$ \cite{ZG18},
iPiano-Net \cite{SL20}, BCSnet \cite{ZHLLZ21} and AMP-Net \cite{ZLLWZ21}, are chosen for comparison.

The experiments use the BSDS500 dataset \cite{AMFM11} which contains 500 color images.
The dataset is divided into a training set of 200 images, a validation set of 100 images, and a test set of 200 images.
170208 image patches of size 33$\times$33 (hence $n$=1089) are randomly extracted from the training set,
and the Adam optimizer \cite{KB14} with a learning rate of 0.001 and a batch size of 64 (i.e., $N$=64) is used to train the network,
with 100 epochs for each compression ratio.
The experiments are conducted on a deep learning workstation with an Intel i7-12700 CPU and GTX2080 GPU.
For testing, two popular benchmark datasets, Set11 \cite{KLTKA16} and BSD68 \cite{MFTM01}, are used.
Peak signal-to-noise ratio (PSNR) and structural similarity (SSIM) \cite{WBSS04} are chosen to evaluate the performance.

Initialize the sampling matrix $\mA$ by randomly selecting proper number of rows from an orthonormal matrix.
Considering the trade-off between network complexity and reconstruction performance,
the default number of DRMs of PIPO-Net is set to $K=9$,
while traditional iterative optimization algorithms usually require $K>100$.

Table~\ref{Comparisons-set11} summarizes the PSNR/SSIM results achieved by the seven methods
with different compression ratios on dataset Set11.
The GPU computing time and the parameter number of each method are also compared.
It can be clearly seen that the proposed PIPO-Net achieves the higher reconstruction accuracy
with acceptable reconstruction speed for different compression ratio cases.

Table~\ref{Comparisons-BSD68} reveals similar conclusions and verifies the superiority of PIPO-Net.
Particularly at low compression ratios of 10\% and 1\%,
PIPO-Net gets better images with 1.05 dB and 0.54 dB higher PSNR values compared to that of AMP-Net, respectively.

\begin{landscape}
\begin{figure}[htb!]
\centering
\subfigure[Ground truth]{
\includegraphics[width=5cm]{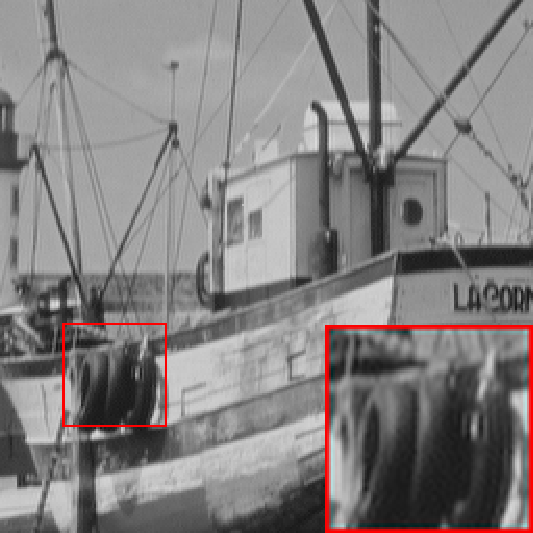}}
\subfigure[ReconNet (25.91/0.7358)]{
\includegraphics[width=5cm]{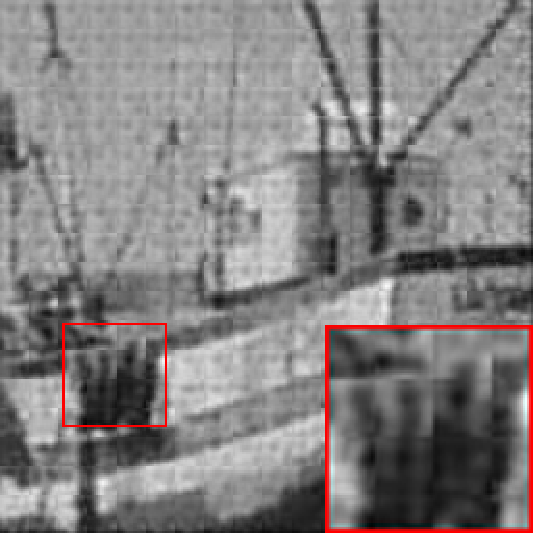}}
\subfigure[CSNet (34.51/0.9322)]{
\includegraphics[width=5cm]{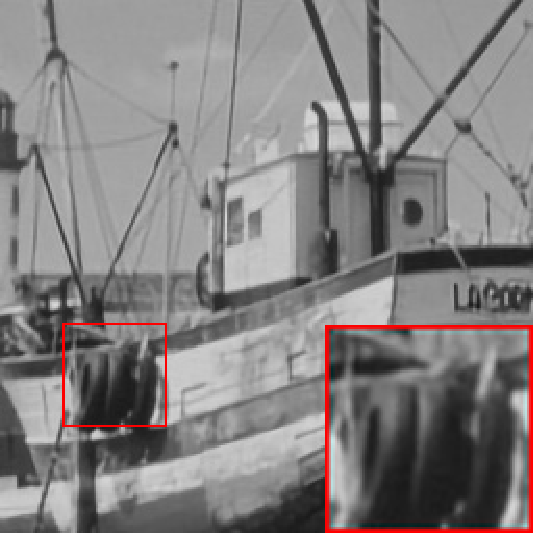}}
\subfigure[ISTA-Net$^+$ (33.69/0.9344)]{
\includegraphics[width=5cm]{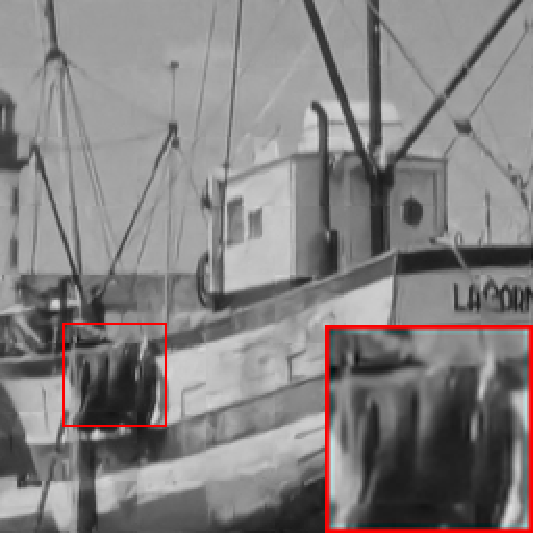}}\\
\subfigure[iPiano-Net (35.37/0.9543)]{
\includegraphics[width=5cm]{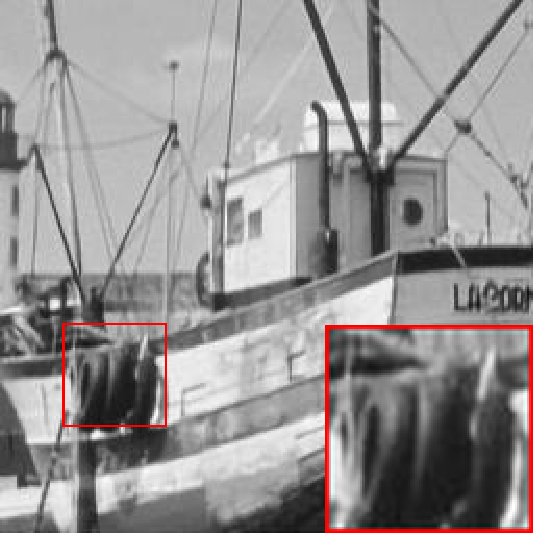}}
\subfigure[BCSnet (35.83/0.9489)]{
\includegraphics[width=5cm]{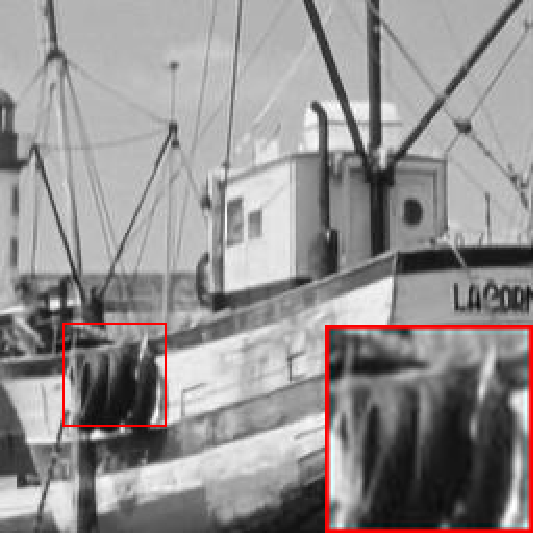}}
\subfigure[AMP-Net (35.56/0.9520)]{
\includegraphics[width=5cm]{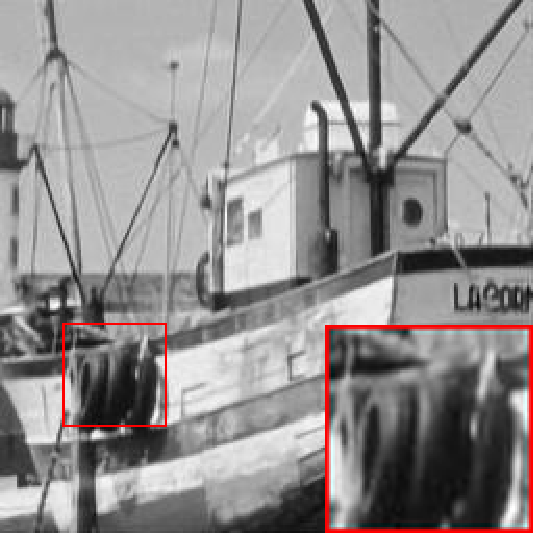}}
\subfigure[PIPO-Net ({\bf 36.01/0.9567})]{
\includegraphics[width=5cm]{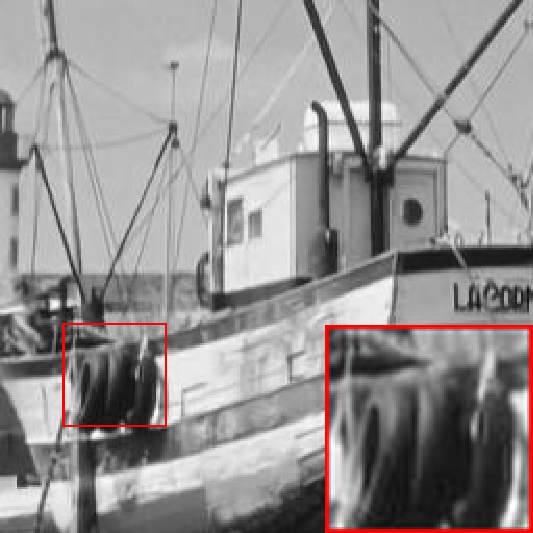}}
\caption{Comparisons of reconstructed boat images by seven methods with 25\% compression ratio.}
\label{boat}
\end{figure}
\end{landscape}

\begin{landscape}
\begin{figure}[htb!]
\centering
\subfigure[Ground truth]{
\includegraphics[width=5cm]{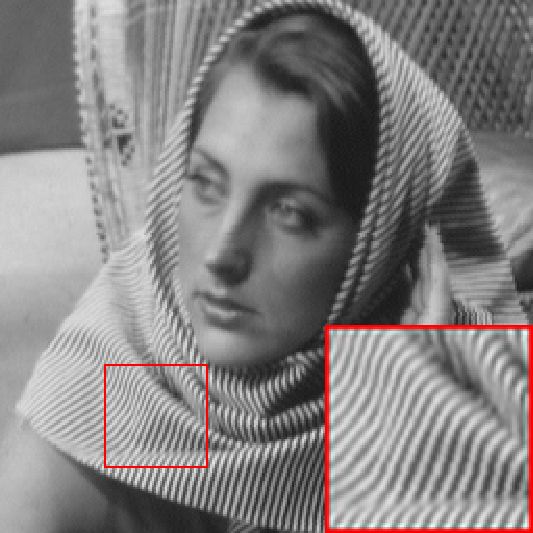}}
\subfigure[ReconNet (19.22/0.6224)]{
\includegraphics[width=5cm]{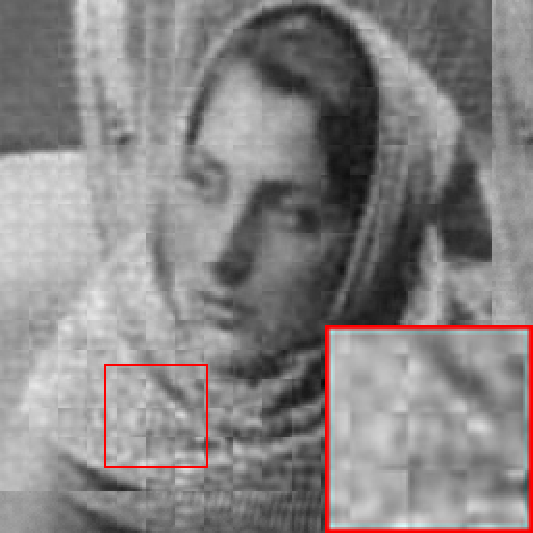}}
\subfigure[CSNet (24.39/0.7390)]{
\includegraphics[width=5cm]{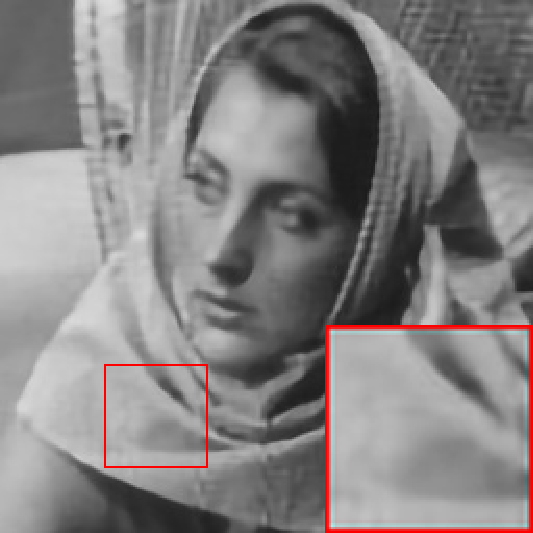}}
\subfigure[ISTA-Net$^+$ (23.53/0.6950)]{
\includegraphics[width=5cm]{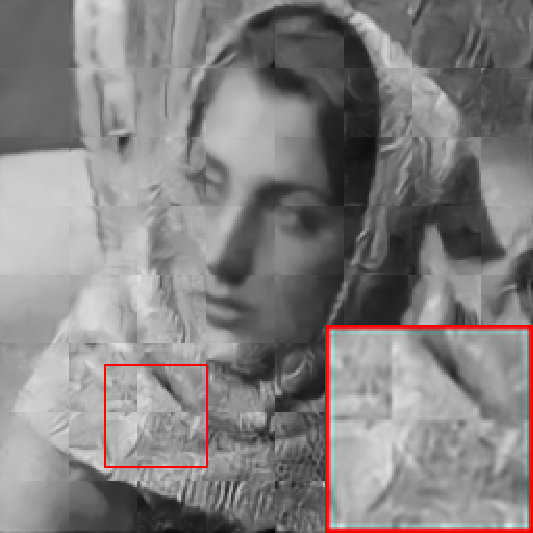}}\\
\subfigure[iPiano-Net (24.75/0.7392)]{
\includegraphics[width=5cm]{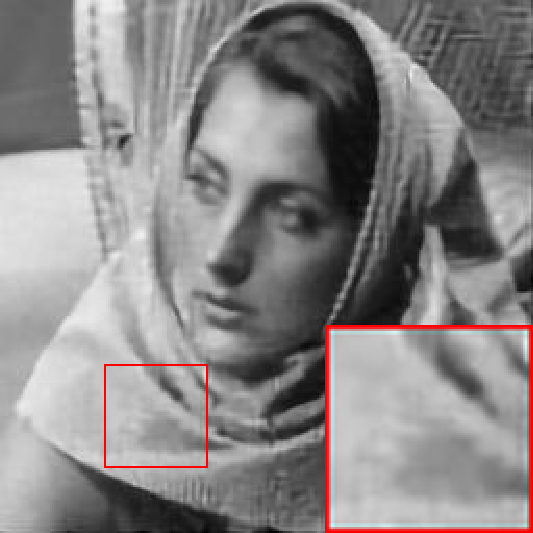}}
\subfigure[BCSnet (25.05/0.7792)]{
\includegraphics[width=5cm]{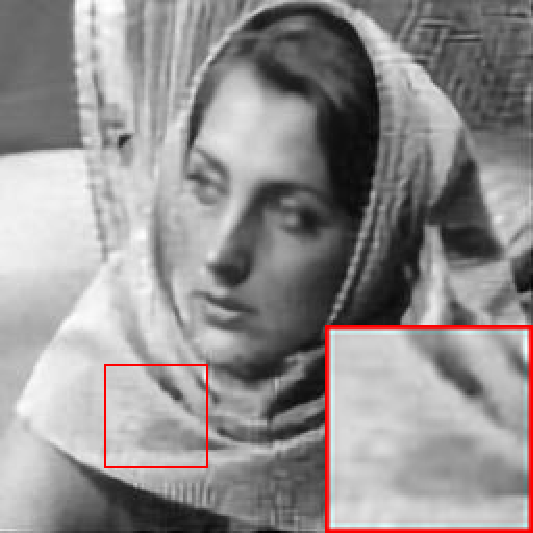}}
\subfigure[AMP-Net (24.84/0.7415)]{
\includegraphics[width=5cm]{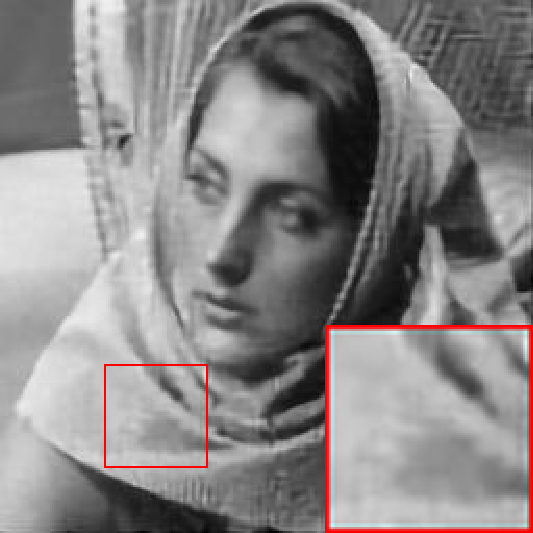}}
\subfigure[PIPO-Net ({\bf 26.82/0.7871})]{
\includegraphics[width=5cm]{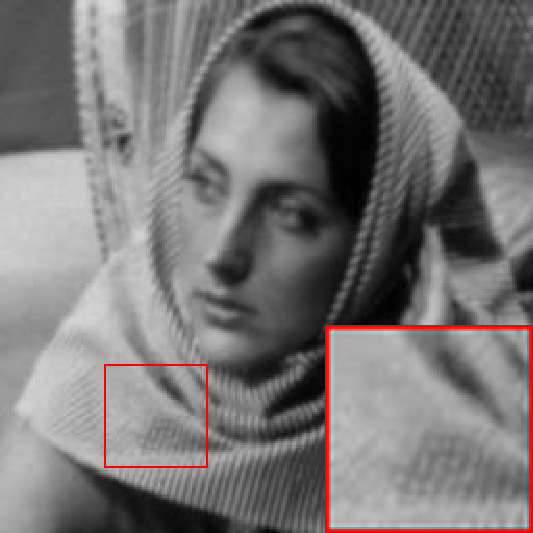}}
\caption{Comparisons of reconstructed Barbara images by seven methods with 10\% compression ratio.}
\label{barbara}
\end{figure}
\end{landscape}

Fig.~\ref{boat} and Fig.~\ref{barbara} show the visual effects together with the PSNR/SSIM results of
reconstructed boat images with 25\% compression ratio
and reconstructed Barbara images with 10\% compression ratio, respectively.
These results highlight the superior performance of PIPO-Net.

\begin{landscape}
\begin{table}[htb!]\caption{Comparisons of PIPO-Net and other six methods on Set11}\label{Comparisons-set11}.
\begin{center}
\begin{tabular}{c||c|c|c|c|c|c|c}
\hline
\hline Method       &50\%          &30\%          &25\%          &10\%          &1\%           &GPU time (s)   &Parameter number (MB) \\
\hline
\hline ReconNet     &31.50/0.9046  &28.14/0.8748  &25.60/0.7613  &24.06/0.7782  &20.09/0.5201  &0.004  &1.22   \\
\hline CSNet        &38.17/0.9631  &34.34/0.9211  &33.56/0.9189  &28.37/0.8169  &20.81/0.5633  &0.026  &2.93   \\
\hline ISTA-Net$^+$ &38.73/0.9685  &34.86/0.9431  &32.57/0.9176  &28.34/0.7341  &17.34/0.4403  &0.012  &2.95   \\
\hline iPiano-Net   &38.88/0.9702  &34.82/0.9441  &33.55/0.9472  &28.12/0.8375  &19.16/0.5122  &0.024  &2.73   \\
\hline BCSnet       &39.57/0.9720  &35.63/0.9495  &34.20/0.9408  &29.42/0.8673  &20.81/0.5427  &0.023  &2.70   \\
\hline AMP-Net      &40.32/0.9804  &36.12/{\bf 0.9602}  &34.64/0.9488  &29.35/0.8779  &20.21/0.5581  &0.027  &3.08   \\
\hline PIPO-Net     &{\bf 40.43/0.9831}  &{\bf 36.32}/0.9598  &{\bf 34.68/0.9500}  &{\bf 29.77/0.9058}  &{\bf 22.06/0.6721}  &0.019  &2.77   \\
\hline\hline
\end{tabular}
\end{center}
\end{table}
\end{landscape}

\begin{table}[htb!]\caption{Comparisons of PIPO-Net and other six methods on BSD68}\label{Comparisons-BSD68}.
\begin{center}
\begin{tabular}{c||c|c|c|c|c}
\hline
\hline Method       &50\%          &30\%          &25\%          &10\%          &1\%                 \\
\hline
\hline ReconNet     &35.93/0.9631  &31.26/0.9024  &28.14/0.8963  &24.06/0.7740  &18.72/0.5418        \\
\hline CSNet        &36.52/0.9798  &31.94/0.9251  &30.91/0.9067  &27.76/0.8513  &21.03/0.5402        \\
\hline ISTA-Net$^+$ &34.92/0.9510  &30.77/0.8901  &29.64/0.8638  &26.25/0.7622  &20.46/0.5211        \\
\hline iPiano-Net   &34.94/0.9519  &31.33/0.8982  &30.52/0.8831  &28.34/0.7442  &20.67/0.5117        \\
\hline BCSnet       &36.69/0.9592  &32.76/0.9341  &31.29/0.9046  &28.98/0.8715  &20.31/0.5319        \\
\hline AMP-Net      &{\bf 37.63/0.9850}  &33.15/0.9352  &32.01/0.9130  &29.40/0.8779  &21.82/0.5571  \\
\hline PIPO-Net     &37.20/0.9801  &{\bf 33.30/0.9384}  &{\bf 32.32/0.9307}  &{\bf 30.45/0.9002}  &{\bf 22.36/0.5976}  \\
\hline\hline
\end{tabular}
\end{center}
\end{table}

To better understand the significance of each part of PIPO-Net, several ablation experiments are carried out on Set11.
The first group of ablation experiments is conducted with compression ratio of 25\%
to verify the importance of the independent update processes of parameters,
involving the penalty parameter $\rho$ and the multiplier vector $\vlambda$.
As can be observed in Table~\ref{Ablation-sharing} that when both tested parameters are optimized independently
without sharing information with each other, the network achieves the best result as indicated in the last row of Table~\ref{Ablation-sharing}.

\begin{table}[htb]\caption{Average PSNR performance of PIPO-Net under different independent update cases of parameters}\label{Ablation-sharing}.
\begin{center}
\begin{tabular}{c|c|c}
\hline
\hline Independent update of $\rho$ &Independent update of $\vlambda$ &PSNR    \\
\hline
\hline $\times$           &$\times$                         &28.77   \\
\hline $\times$           &$\surd$                          &32.37   \\
\hline $\surd$            &$\times$                         &31.59   \\
\hline $\surd$            &$\surd$                          &34.68   \\
\hline\hline
\end{tabular}
\end{center}
\end{table}

Table~\ref{Ablation-HFC} summarizes the results of the second group of ablation experiments
to assess the performance of the MSS and the HFC block with different compression ratios.
The highest PSNR values shown in the last row fully verify the effectiveness of these two processes.

\begin{table}[htb]\caption{Ablation experiments on assessing the performance of MSS and HFC}\label{Ablation-HFC}.
\begin{center}
\begin{tabular}{c|c|c|c|c}
\hline
\hline MSS           &HFC       &25\%   &10\%   &1\%    \\
\hline
\hline $\times$      &$\times$  &30.63  &22.05  &14.91   \\
\hline $\times$      &$\surd$   &33.83  &28.99  &21.33   \\
\hline $\surd$       &$\times$  &33.90  &29.23  &21.48   \\
\hline $\surd$       &$\surd$   &34.68  &29.77  &22.06   \\
\hline\hline
\end{tabular}
\end{center}
\end{table}

The last group of ablation experiments is carried out to assess the performance of different loss functions.
The results in Table~\ref{Ablation-lossF} show the superiority of the proposed loss function.

\begin{table}[htb]\caption{Ablation experiments on assessing the performance of different loss functions}\label{Ablation-lossF}.
\begin{center}
\begin{tabular}{c|c|c|c}
\hline
\hline Loss function             &25\%   &10\%   &1\%   \\
\hline
\hline $\calL_{\text{MSE}}$      &34.55  &28.55  &20.53 \\
\hline $\calL_{\text{total}}$    &34.68  &29.77  &22.06 \\
\hline\hline
\end{tabular}
\end{center}
\end{table}

\section{Conclusion}\label{section4}
In this paper, a novel deep unfolding architecture called PIPO-Net
has been proposed for reconstructing images by optimizing the penalty-based CS problems.
All the learnable parameters in one module are updated independently from those of other modules.
The MSS and the HFC blocks are developed in PIPO-Net to enhance its performance.
Experiments have been carried out to verify the superiority of the proposed PIPO-Net.

\vspace{0.25cm}\noindent
{\bf Acknowledgement}
\vspace{0.25cm}

This work was supported in part by the Natural Science Foundation of China under Grants 61571174 and 61801159,
and in part by the China-Montenegro Bilateral Science \& Technology Cooperation Project.


\end{document}